\documentclass[12pt]{article}
\usepackage{psfig}

\begin{document}

\title{Necessary Conditions on Realizable
Two-Point Correlation Functions of Random Media}
\author {S. Torquato \\ Department of Chemistry, \\
Program in Applied and Computational Mathematics, PRISM,\\ 
and Princeton Center for Theoretical Physics\\
Princeton Univeristy, Princeton, New Jersey 08540 }
\maketitle

\begin{abstract}
A fascinating inverse problem that has been receiving considerable
attention is the construction of realizations of random two-phase heterogeneous media
with a target two-point correlation function.
However, not every hypothetical two-point correlation function
corresponds to a realizable two-phase medium. Here we collect
all of the known necessary conditions on the two-point correlation
functions scattered throughout a diverse literature and derive a 
new but simple positivity condition. We apply the necessary conditions
to test the realizability of certain classes
of proposed correlation functions.
\end{abstract}
\maketitle

\section{Introduction}

Random two-phase  heterogeneous media abound in synthetic products and nature.
Examples include composite materials, colloidal dispersions, gels, foams,
wood, geologic media, and animal and plant tissue.\cite{Ch79,Ru89,To02a,Sa03,Qu03}
The effective transport, mechanical and electromagnetic properties of
such heteroegeneous materials are known to depend on
correlation functions that statistically characterize the microstructure.\cite{To02a}
It has recently been suggested that microstructure reconstruction problems
can be posed as  optimization problems.\cite{Ri97a,Ye98a}
A set of {\em target} correlation functions are prescribed
based upon  experiments or theoretical models. Starting from 
some initial realization of the random two-phase medium, the 
reconstruction method proceeds to find 
a realization  by evolving the microstructure such that the calculated correlation
functions best match the target functions. This intriguing
inverse problem is solved by minimizing an {\it error} based upon the distance between the
target and calculated correlation functions. The two-phase medium can be a 
dispersion of particles in some matrix (liquid or solid) \cite{Ri97a} or, more 
generally, a digitized image of a two-phase material.\cite{Ye98a}

An effective reconstruction procedure enables one to generate accurate structures at will,
and subsequent analysis can be performed on the image to obtain
desired macroscopic properties (e.g., transport, electromagnetic and mechanical
properties) of the media. This becomes especially useful in generating
three-dimensional structures from planar information when three-dimensional imaging
techniques are not available: a ``poor man's'' tomography experiment.\cite{Ye98b,To02a}

Interestingly, the same procedure has been used to ``construct"
realizations of two-phase media from a hypothetical target correlation
function.\cite{Ye98a,Cu99,Sh01,To02a}
In this mode, the procedure is referred to as a {\it construction} algorithm. 
There are mnay different types of statistical descriptors of two-phase
media, \cite{To02a} but the most basic one is the two-point correlation function,
which gives the probability of finding two points in one of the
phases (see definition below) and is obtainable from small-angle X-ray scattering.\cite{De49}
The construction algorithm can be employed to determine if a 
prescribed two-point correlation function is in fact realizable.
If such a two-point correlation function is realizable, then the
procedure could be used to categorize classes of random microstructures,
which would be a valuable accomplishment.
However, not every hypothetical two-point correlation function
corresponds to a realizable two-phase medium \cite{To02a}. Therefore, it is
of great fundamental and practical importance to determine the
necessary conditions that realizable two-point correlation
functions must possess.\cite{To99c,To02a} We note in passing
that this question is closely related to realizability issues
of pair-correlation functions of many-particle systems.\cite{To02c,Cr03,Cos04} 

One aim of this paper is to gather all of the known necessary conditions
on the two-point correlation function of two-phase
random media, also known as ``random closed sets" in
the field of stochastic geometry.\cite{St95,Ko03} Some of these conditions are well-known
in the physical sciences  literature, but others are more arcane
and are contained in obscure mathematical technical reports
and/or proceedings. We also derive a 
new but simple positivity condition on the two-point correlation function.
We consider  some illustrative examples
of proposed correlation functions and test whether they can correspond 
to realizable two-phase random media.

\section{Necessary Conditions}

Here we collect all of the known necessary conditions on the two-point correlation
function of random media that are scattered throughout a diverse literature. We also derive a
new positivity condition.

Each realization $\omega$ of the two-phase random medium occupies the region of
$d$-dimensional Euclidean space ${\cal V}\in{\Re}^d$ of volume $V$
that is partitioned into two {\it random} sets  or {\it phases} whose
interiors are disjoint: phase~1,
a region ${\cal V}_1(\omega)$ of volume fraction $\phi_1$, and phase~2, a region 
${\cal V}_2(\omega)$ of volume fraction $\phi_2$. For a given realization 
$\omega$, the {\it indicator function} ${\cal I}^{(i)}({\bf x};\omega)$ for 
phase~$i$ at any position vector $\bf x \in {\cal V}$ is defined by
\begin{equation}
{\cal I}^{(i)}({\bf x};\omega) = \cases{ 1, &  $ \mbox{~if~ ${\bf x} \in {\cal V}_i(\omega)$} $,
\cr
0, &  $ \mbox{~otherwise} $, \cr}
\label{char}
\end{equation}
Thus, a two-phase random medium is described by a binary stochastic process 
$\{ {\cal I}^{(i)}(\bf x) : \bf x \in \Re^d\}$. For statistically homogeneous but
anisotropic media, 
the first two correlation functions are given by\cite{To02a}
\begin{equation}
S_1^{(i)}({\bf x})=\langle {\cal I}^{(i)}({\bf x}) \rangle = \phi_i
\end{equation}
and
\begin{equation}
S_2^{(i)}({\bf r})=\langle {\cal I}^{(i)}({\bf x}_1) {\cal I}^{(i)}({\bf x}_2) \rangle,
\end{equation}
where $i=1$ or $2$, angular brackets denote an ensemble average, 
${\bf r}={\bf x}_1-{\bf x}_2$, and the symbol $\omega$ is henceforth dropped for brevity.
(The generalization to $n$-point correlation functions for $n\ge 1$ is straightforward.\cite{To02a})
Clearly, $\phi_i$ lies in the closed interval $[0,1]$ and $\phi_1+\phi_2=1$.
The two-point or autocorrelation function $S_2^{(i)}({\bf r})$ for statistically 
homogeneous media gives the probability of finding the end points of 
a vector $\bf r$ in phase $i$. Debye and Bueche \cite{De49} showed that the two-point 
correlation function of a porous medium can be obtained experimentally
via small X-ray scattering. Note that the two-point function for phase 2 is
simply related to the corresponding function for phase 1 via the expression
\begin{equation}
S_2^{(2)}({\bf r})=S_2^{(1)}({\bf r})-2\phi_1+1,
\end{equation}
and thus the {\it autocovariance} function
\begin{equation}
\chi({\bf r}) \equiv S_2^{(1)}({\bf r})-\phi_1^2=S_2^{(2)}({\bf r})-\phi_2^2
\label{covariance}
\end{equation}
for phase 1 is equal to that for phase 2. Generally, for $\bf r=0$, 
\begin{equation}
S_2^{(i)}({\bf 0})=\phi_i,
\label{0}
\end{equation}
and in the absence of no {\it long-range} order, 
\begin{equation}
\lim_{|{\bf r}| \rightarrow \infty} S_2^{(i)}({\bf r}) \rightarrow \phi_i^2.
\label{infty}
\end{equation}

An important  necessary condition for the existence of a two-point correlation 
function $S_2^{(i)}({\bf r})$ for a two-phase statistically homogeneous medium 
$d$ dimensions is that the $d$-dimensional
Fourier transform of autocovariance $\chi({\bf r})$, denoted
by ${\tilde \chi}({\bf k})$, must be nonnegative for all wave vectors,\cite{To02a} i.e.,
\begin{equation}
{\tilde \chi}({\bf k})=\int_{\Re^d} \chi({\bf r}) e^{\displaystyle -i{\bf k}\cdot{\bf r}}\ 
d{\bf r} \ge 0, \quad \mbox{for all}\;\; {\bf k},
\label{th-wiener}
\end{equation}
where $\chi({\bf r})$ is given by (\ref{covariance}). Physically, this
nonnegativity condition arises because ${\tilde \chi}({\bf k})$ is proportional
to the scattered intensity, which must be positive.\cite{To02a}
This is sometimes called the Wiener--Khintchine condition,\cite{To02a} which
is necessary but not sufficient for the class $B$ of correlation functions
that come from binary stochastic processes $\{ {\cal I}^{(i)}(\bf x) : \bf x \in \Re^d\}$.
The Wiener--Khintchine condition is easily proved by exploiting
a well-known theorem that states any continuous  function $\chi(\bf r)$ must be positive
semi-definite  in the sense that for any finite number
of spatial locations ${\bf r}_1,{\bf r}_2,\ldots, {\bf r}_m$ in $\Re^d$
and arbitrary real numbers $a_1,a_2,\ldots,a_m$,
\begin{equation}
\sum_{i=1}^m \sum_{j=1}^m a_ia_j \chi({\bf r}_i-{\bf r}_j) \ge 0
\label{chi-pos}
\end{equation}
if and only if it has a nonnegative Fourier transform ${\tilde \chi}({\bf k})$.
Note that this property does not prevent $\chi({\bf r})$ from being pointwise negative 
for certain $\bf r$. Importantly, whereas the real-space condition is difficult to check,
the spectral version (\ref{th-wiener}) is straightforward to test.
It is noteworthy that if the medium in $d$ dimensions is both statistically homogeneous and isotropic,
then the one-, two-, $\cdots$ and $d$-dimensional Fourier transforms 
of $\chi({\bf r})$ must all be nonnegative.\cite{To99c} This is a consequence
of the fact that $\chi({\bf r})$ for such a random medium is an invariant in 
any $m$-dimensional subspace, where $m=1,2,\ldots,d-1$.

The task of determining the necessary and sufficient conditions
that $B$ must possess is very complex. It has been shown that 
autocovariance functions in $B$ must not only meet the condition of (\ref{th-wiener})
but another condition on ``corner-positive'' matrices.\cite{To02a}
Since little is known about corner-positive matrices,
this theorem is very difficult to apply in practice.
Thus, a meaningful characterization of $B$ remains
an open and interesting problem, especially in the
context of $d$-dimensional two-phase random media.

No attempt will be made to address the complete characterization
of $B$ here but instead we summarize the known 
necessary conditions, in addition to condition (\ref{th-wiener}),
that characterize $B$, most of which are described in Ref. (3).
The two-point correlation function must satisfy the bounds
\begin{equation}
0 \le S_2^{(i)}({\bf r}) \le \phi_i \qquad \mbox{for all} \;\; {\bf r}.
\label{chi-0}
\end{equation}
The lower bound states that $S_2^{(i)}(\bf r)$ must be nonnegative
for all $\bf r$, but we show below that either $S_2^{(1)}(\bf r)$ or $S_2^{(2)}(\bf r)$  must strictly be positive
for $\phi_i\neq 1/2$. The corresponding bounds on the autocovariance function is given by \cite{To02a}
\begin{equation}
- \min(\phi_1^2,\phi_2^2) \le \chi({\bf r}) \le \phi_1\phi_2 \qquad \mbox{for all} \;\; {\bf r}.
\label{chi-1}
\end{equation}
Another consequence of the binary nature of the process in
the case of statistically homogeneous and isotropic media, i.e., 
when $S_2^{(i)}(\bf r)$ only depends on the distance $r \equiv |\bf r|$, is that
its derivative at $r=0$ is strictly negative or
\begin{equation}
\frac{d S^{(i)}_2}{dr} \Bigg\vert_{r=0} = \frac{d \chi}{dr} \Bigg\vert_{r=0} 
<0
\qquad \mbox{for all} \;\; 0 <\phi_i <1.
\label{chi-slope}
\end{equation}
This is a consequence of the fact that slope at $r=0$ is proportional to the
negative of the specific surface.\cite{To02a} This means that 
$S_2^{(i)}(r)$ has a cusp at the origin, implying that the two-point
function is nonanalytic at the origin. It is a property of binary processes that if $\chi_1({\bf r})$ and
$\chi_2({\bf r})$ are in $B$, then $\chi_1({\bf r}) \cdot \chi_2({\bf r})
\in B$
and $\alpha \chi_1({\bf r}) +(1-\alpha)\chi_2({\bf r}) \in B$ for
every $\alpha \in [0,1]$. This was proved by Shepp\cite{Sh63}
in one dimension, but the proof should extend trivially to $d$
dimensions.

A little known necessary condition for statistically homogeneous media 
is the so-called ``triangular inequality" first derived by Shepp \cite{Sh63} 
and later rediscovered by Matheron:\cite{Ma93}
\begin{equation}
S_2^{(i)}({\bf r})\ge S_2^{(i)}({\bf s})+S_2^{(i)}({\bf t}) -\phi_i,
\label{tri}
\end{equation}
where ${\bf r}={\bf t}-{\bf s}$. The derivation of the
triangular inequality (\ref{tri}) is straightforward. Following
Shepp, we introduce the random variable
\begin{equation}
Y^{(i)}({\bf x})=2{\cal I}^{(i)}({\bf x})-1 = 
\cases{ 1, &  $ \mbox{~if~ ${\bf x} \in
{\cal V}_i$} $,
\cr
-1, &  $ \mbox{~otherwise} $, \cr}.
\label{char2}
\end{equation}
The mean of $Y^{(i)}({\bf x})$ is $\langle Y^{(i)}({\bf x})\rangle =
2\phi_i-1$, which is equal to zero if $\phi_1=\phi_2=1/2$.
Observe that $Y^{(i)}({\bf x}_1)-
Y^{(i)}({\bf x}_2)+Y^{(i)}({\bf x}_3)$ is an odd number (either $-3, -1,1$ or $3$) and 
therefore
\begin{equation}
\langle [Y^{(i)}({\bf x}_1)- Y^{(i)}({\bf x}_2)+Y^{(i)}({\bf x}_3)]^2
\rangle \ge 1.
\end{equation}
Using the fact that $\langle Y^{(i)}({\bf x}_1)Y^{(i)}({\bf x}_2) \rangle=
4S_2^{(i)}({\bf x}_1-{\bf x}_2)-4\phi_i+1$, where we have invoked statistical
homogeneity, we immediately obtain
the triangular inequality (\ref{tri}). 

Note that if the autocovariance $\chi(r)$
of a statistically homogeneous and isotropic medium
is monotonically decreasing, nonnegative and convex (i.e., $d^2\chi(r)/dr^2 \ge 0$),
then it satisfies the triangular inequality (\ref{tri}). \cite{Ma95}
The triangular inequality implies a number of pointwise conditions
on the two-point correlation function. For example, for statistically homogeneous and isotropic media,
the triangular inequality implies condition (\ref{chi-slope}), the fact that the 
steepest descent of  the two-point correlation function occurs at the origin,\cite{Sh63} i.e.,
\begin{equation}
|S_2^{(i)}(0)| \ge |S_2^{(i)}(r)|  \qquad \mbox{for all} \;\; {r},
\end{equation}
and the fact that $S_2^{(i)}(r)$ must convex at the origin,\cite{Ma95} i.e.,
\begin{equation}
\frac{d^2 S_2^{(i)}}{dr^2}  \Bigg\vert_{r=0}=\frac{d^2 \chi}{dr^2} \Bigg\vert_{r=0}\ge 0.
\label{convex}
\end{equation}
From the ``stochastic continuity" theorem for general stochastic
processes,\cite{Pr81} it follows that if $S_2^{(i)}(r)$ is continuous
at $r=0$, then it is continuous for all $r$. This continuity
condition can also be proved using the triangular inequality.
Note that $S_2^{(i)}(r)$ can be discontinuous at the origin
if the specific surface $s$ is infinitely large.

The triangular inequality is actually
a special case of the more general condition \cite{Sh63}
\begin{equation}
\sum_{i=1}^m \sum_{j=1}^m \epsilon_i\epsilon_j \chi({\bf r}_i-{\bf r}_j) \ge 1, \qquad \epsilon_i \pm 1, \quad i=1,\ldots, m, 
\quad m ~\ \mbox{odd}.
\label{chi-pos-2}
\end{equation}
This necessary condition is much stronger than (\ref{chi-pos}), implying that there
are other necessary conditions beyond the ones identified so far. However, 
the condition (\ref{chi-pos-2}) is difficult to check in practice because
it does not have a simple spectral analog in contrast to (\ref{chi-pos}) [cf. 
(\ref{th-wiener})]. Note that the integers $\epsilon_i=\pm 1$ in (\ref{chi-pos-2})
can be replaced with general integers, which would lead to an even 
more general condition on $\chi({\bf r})$.

Here we report a new simple consequence of the lower bound of expression (\ref{chi-1}).
Because the autocovariance is the same for phase 1 and phase 2, then it immediately
follows from the lower bound of (\ref{chi-1}) that
\begin{equation}
S_2^{(i)}({\bf r}) \ge \max(0, 2\phi_i-1)  \qquad \mbox{for all} \;\; {\bf r}.
\label{positive}
\end{equation}
Thus, for $\phi_i >1/2$, $S_2^{(i)}({\bf r})$ is strictly positive such
that it must be greater than $2\phi_i -1$. Interestingly, the lower bound of (\ref{chi-1})
for the autocovariance $\chi(\bf r)$, first obtained in Ref. 3, was derived from the trivial
pointwise nonnegativity condition $S_2^{(i)}({\bf r}) \ge 0$. However, the consequences
of going back to the two-point correlation function $S_2^{(i)}({\bf r})$
were heretofore not examined. The nontrivial positivity condition (\ref{positive})
arises because the statistics of phase 1 are not independent of the statistics
of phase 2. Since $\phi_i^2$ is the large-distance asymptotic limit of $S_2^{(i)}({\bf r})$,
its globally minimum value or, more precisely, its infimum (greatest lower bound)  must
be less than or equal to $\phi_i^2$. (Technically, one must consider the infimum 
and not the minimum because the minimum may not actually be achieved, e.g., a monotonically
decreasing function that only asymptotically approaches its minimum value
of $\phi_i^2$.) Clearly, the lower bound (\ref{positive}) holds for the 
infimum of $S_2^{(i)}({\bf r})$, which will be denoted by $\inf[S_2^{(i)}({\bf r})]$.
In summary, the infimum of any two-point correlation function of a statistically homogeneous 
medium must satisfy the inequalities
\begin{equation}
\max(0, 2\phi_i-1) \le \inf[S_2^{(i)}(r)] \le \phi_i^2
\label{min}
\end{equation}
(see Figure 1).

\begin{figure}[ht]
\centerline{\psfig{file=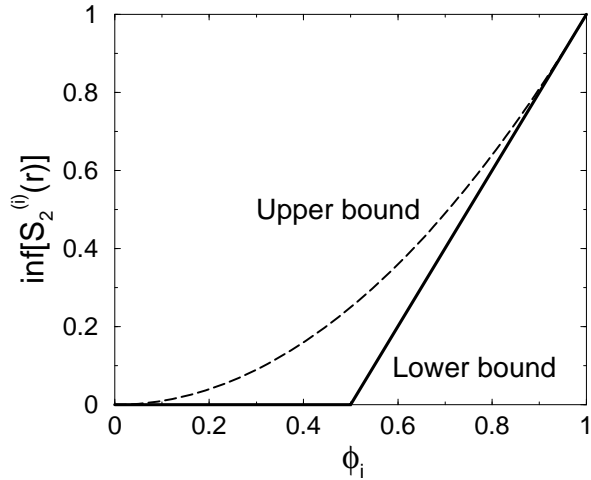,width=3in}}
\caption{Graphs of the upper bound (dashed curve) and lower
bound (solid lines) of (\ref{min}) on the infimum of $S_2^{(i)}(r)$
for a  statistically homogeneous medium.}
\label{pos}
\end{figure}



\section{Illustrative Examples}

It is convenient to introduce the scaled autocovariance function $f(\bf r)$ 
defined as 
\begin{equation}
f({\bf r}) \equiv \frac{\chi({\bf r})}{\phi_1\phi_2}=\frac{S_2^{(i)}({\bf
r})-\phi_i^2}{\phi_1\phi_2} \qquad \mbox{for} \;\; 0 \le r < +\infty.
\end{equation}
From (\ref{positive}), we obtain the triangular inequality for $f$ to be
\begin{equation}
f({\bf r})\ge f({\bf s})+f({\bf t}) -1.
\label{tri-2}
\end{equation}
Moreover, the bounds (\ref{chi-1})  become
\begin{equation}
-\min\Big[\frac{\phi_1}{\phi_2}, \frac{\phi_2}{\phi_1}\Big]  \le f({\bf r}) \le 1
\qquad \mbox{for all} \;\; {\bf r}.
\label{positive-2}
\end{equation}

\begin{figure}[ht]
\centerline{\psfig{file=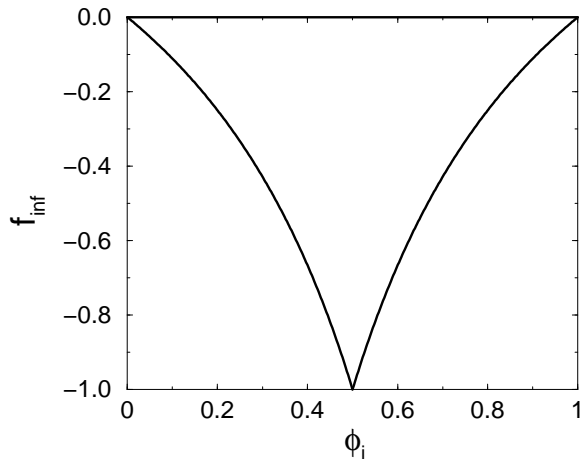,width=3.0in}}
\caption{The bounds (\ref{min-2}) on the  infimum $f_{inf}$ for volume-fraction-independent 
$f(r)$. The quantity $f_{inf}$ must lie within the region delineated by the heavy solid lines.}
\label{fmin}
\end{figure}

Our focus in this paper will be hypothetical continuous functions
$f(r)$ that depend on the distance $r=|\bf r|$  and could potentially 
correspond to statistically homogeneous and isotropic media without long-range
order such that $f(0)=1$ and $f(r)$ tends to zero as $ r \rightarrow \infty$ sufficiently
fast so that the Fourier transform of $\chi(r)=S_2^{(i)}(r)-\phi_i^2$ exists.
The latter two properties of $f(r)$ ensure that $S_2^{(i)}(r)$ obeys its proper
asymptotic limiting behaviors as specified by (\ref{0}) and (\ref{infty}), respectively.
When the scaled autocovariance  $f(r)$ depends only on the magnitude $r=|{\bf r}|$, then
the Fourier transform condition (\ref{th-wiener}) on ${\tilde f}(k)$ can be written in any space
dimension $d$  as\cite{To03a}
\begin{eqnarray}
{\tilde f}(k)&=& \left(2\pi\right)^{\frac{d}{2}}\int_{0}^{\infty}r^{d-1}f(r)
\frac{J_{\left(d/2\right)-1}\!\left(kr\right)}{\left(kr\right)^{\left(d/2\right)-1}}dr \ge 0,
\label{necess}
\end{eqnarray}
where $k=|{\bf k}|$ and $J_{\nu}(x)$ is the Bessel function of order $\nu$.
The bounds (\ref{min}) are equivalent to
\begin{equation}
-\min\Big[\frac{\phi_1}{\phi_2}, \frac{\phi_2}{\phi_1}\Big] \le f_{inf} \le 0,
\label{min-2}
\end{equation}
where $f_{inf}$ is the infimum of $f(r)$.  
Note that when function $f(r)$ is independent of the volume fraction $\phi_1$,
it would correspond, if realizable,  to a two-phase medium with 
{\it phase-inversion symmetry}.\cite{To02a}  
A two-phase  random medium possesses phase-inversion symmetry  if
the geometry of phase 1 at volume fraction $\phi_1$
is statistically identical to that of phase 2 in the system
where the volume fraction of phase 1 is  $\phi_2$ and hence
\begin{equation}
S^{(1)}_{2}({\bf r},\phi_1,\phi_2)=S^{(2)}_{2}({\bf r};\phi_2,\phi_1).
\label{sym-0.5}
\end{equation}
By construction, the upper bound of (\ref{min-2}) is always satisfied.
All of the functions $f(r)$ considered below
are taken to be independent of the volume fraction $\phi_1$ and therefore
any violation of the lower bound of (\ref{min-2}) implies that 
a two-phase statistically homogeneous and isotropic medium
cannot exist for the following volume-fraction intervals
\begin{equation}
0 < \phi_i < \frac{|f_{inf}|}{1+|f_{inf}|} \qquad  \mbox{and} \qquad \frac{1}{1+|f_{inf}|} <\phi_i <1.
\label{intervals}
\end{equation}
Figure \ref{fmin} depicts the bounds (\ref{min-2}) on $f_{inf}$ for $f(r)$
that are independent of volume fraction.

\begin{figure}[ht]
\centerline{\psfig{file=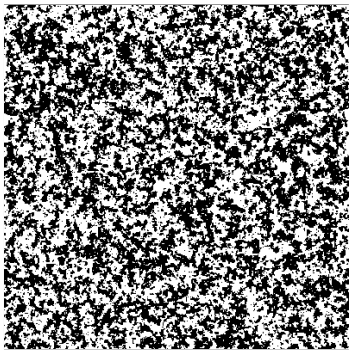,width=1.8in}}
\caption{Construction of a digitized two-dimensional
realization of a ``Debye'' random medium ($400\times400$ pixels).\cite{Ye98a,To02a}
Here the volume fraction $\phi_1=\phi_2=0.5$ and correlation length $a=2$ pixels.}
\label{Debye-medium}
\end{figure}

First we note that for any
$f(r)$ that monotonically decreases in $r$ to its long-range value of zero,
the pointwise nonnegativity condition (\ref{positive-2})
is obeyed for $0\le \phi_i \le 1$. However, as some examples below 
will demonstrate, such an $f(r)$ does not necessarily obey the triangular inequality (\ref{tri-2}).
A natural example of a monotonic scaled autocovariance function $f(r)$ is the simple exponentially decaying
function, i.e.,
\begin{equation}
f(r)= \exp(-r/a),
\label{debye}
\end{equation}
where $a$ is a positive parameter that we call the ``correlation length." This function
was first proposed by Debye and coworkers,\cite{De49,De57} who intuited that it should 
correspond to structures in which one phase consists of ``random shapes and sizes," but presented no proof
that such was the case. The function (\ref{debye}) obeys the necessary
nonnegativity condition (\ref{necess}) on the spectral function ${\tilde f}(k)$
for any $d$ as well as the triangular inequality (\ref{tri-2}). The satisfaction
of these necessary conditions does not ensure that such a correlation is realizable.
However, the aforementioned inverse optimization construction technique \cite{Ye98a,To02a} was applied
to generate a two-dimensional digitized realization corresponding to (\ref{debye}) (see Figure~\ref{Debye-medium}). 
This leads one to believe that (\ref{debye}) is exactly realizable. Indeed, there are specific
two-phase microstructures that achieve the ``Debye'' random-medium function
(\ref{debye}) in the plane. \cite{St95} The function (\ref{debye}) is a special
case of a more general realizable subclass of $B$ given by the {\it completely monotonic
functions},\cite{Sh63} i.e.,
\begin{equation}
f(r)=\int_0^\infty \exp(-\lambda r) \; dF(\lambda),
\end{equation}
where $F(\lambda)$ is a nonnegative bounded measure [bounded and nonincreasing function
on $(0,\infty)$], i.e., $dF \ge 0$ and $\int_0^\infty dF(\lambda)=1$. We see that
if $F= \Theta(\lambda-a^{-1})$, then $dF=\delta(\lambda-a^{-1})$ and (\ref{debye})
is recovered, where $\Theta(x)$ and $\delta(x)$ are the Heaviside and Dirac
delta functions, respectively.
     
\begin{figure}[ht]
\centerline{\psfig{file=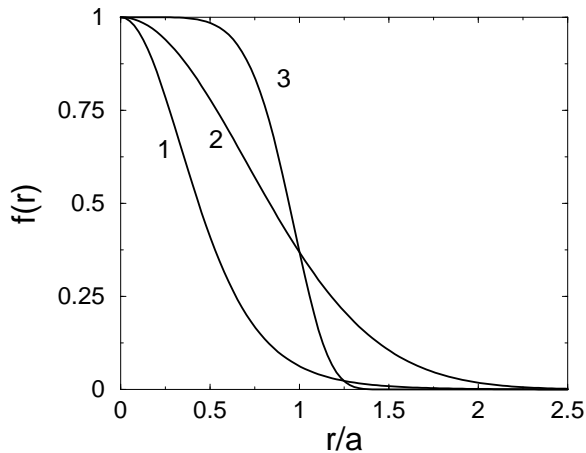,width=3in}}
\caption{Examples of scaled autocovariance functions
that cannot correspond to statistically homogeneous
and isotropic two-phase random media: (1) $f(r)=1/(1+r^2)^4$ for $d \le 5$;
(2) $f(r)=  \exp(-r^2)$ for any $d$; and (3) $f(r)=  \exp(-r^6)$ for any $d$.}
\label{impossible}
\end{figure}

Another natural monotonic scaled autocovariance function  $f(r)$ to consider is the Gaussian function, i.e., 
\begin{equation}
f(r)= \exp[-(r/a)^2].
\label{gaussian}
\end{equation}
Although any such Gaussian function has a nonnegative
spectral function ${\tilde f}(k)$, it cannot correspond to a two-phase
random medium in $\Re^d$ because the slope of $S_2^{(i)}(r)$ at $r=0$ is zero (i.e.,
the specific surface is zero) and therefore violates condition (\ref{chi-slope}) or, more 
generally, the triangular inequality (\ref{tri-2}).  For precisely the same
reasons, the class of monotonic functions 
\begin{equation}
f(r)=  \exp[-(r/a)^\alpha]. \qquad \mbox{for any} \;\; \alpha  > 1
\label{decay1}
\end{equation}
and
\begin{equation}
f(r)= \frac{1}{[1+(r/a)^2]^{\beta-1}} \qquad \mbox{for any} \;\; \beta \ge d
\label{decay2}
\end{equation}
cannot correspond to a two-phase random medium in $d$ dimensions.
These specific examples, some of which
are illustrated in Fig. \ref{impossible}, show that the nonnegativity condition (\ref{necess}) and triangular inequality
(\ref{tri}) are independent necessary conditions.

The final monotone function that we test is the 
simple linear function
\begin{equation}
f(r)= \cases{ 1 -r/a, &  $ \mbox{~if~ $r \le a$} $,
\cr
0, &  $ \mbox{~otherwise} $, \cr}
\label{linear}
\end{equation}
Shepp \cite{Sh63} proved that such a scaled autocovariance is
realizable by a statistically homogeneous two-phase medium
in one dimension. However, this autocovariance is
not realizable in higher dimensions because its spectral
function ${\tilde f}(k)$ can take on negative values
for certain values of $k$. It is noteworthy
that it has been shown that for any positive definite
$f(r)$ in one dimension, the function $2 \arcsin(f)/\pi$ 
as well as $8\pi^{-2} \sum_k (2k+1)^2 f((2k+1)r)$ are in $B$. \cite{Sh63}

\begin{figure}[ht]
\centerline{\psfig{file=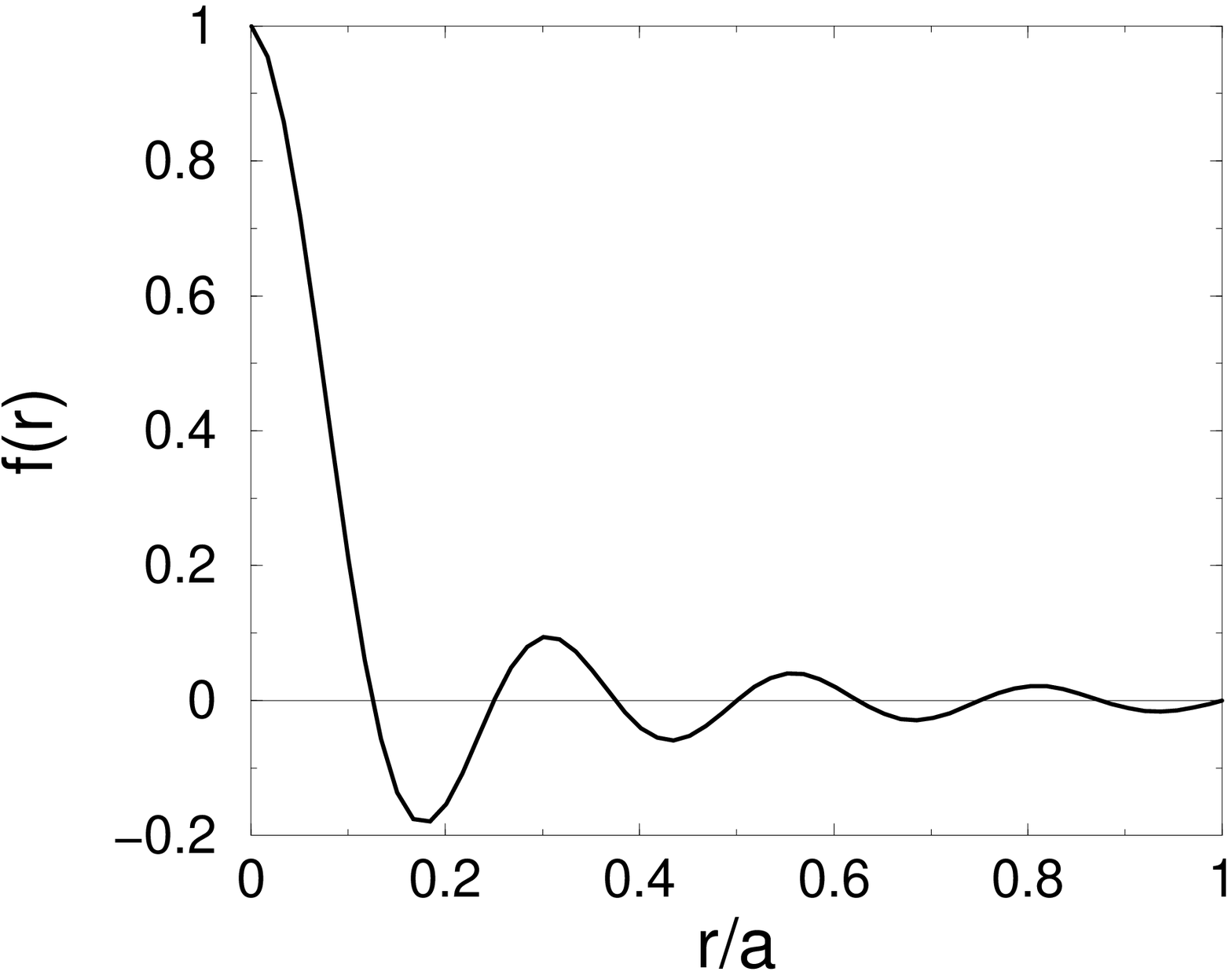,width=3.0in}}
\caption{The damped sinusoidal function (\ref{E10}) with $qa=8\pi$.}
\label{damped-sine}
\end{figure}

\begin{figure}
\centerline{\psfig{file=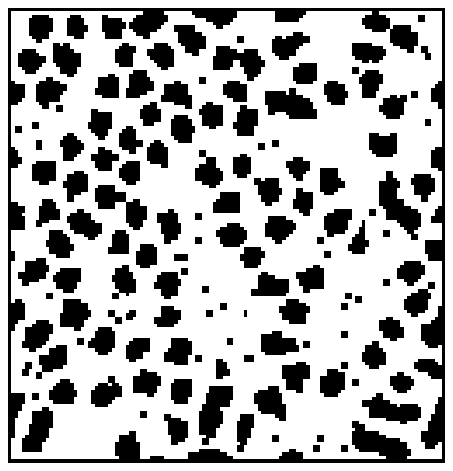,width=1.8in}
\hspace{0.15in}\psfig{file=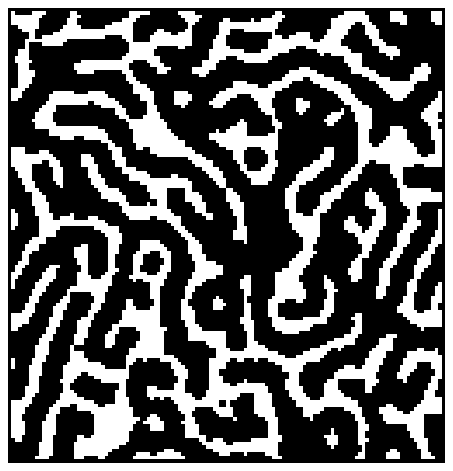,width=1.8in}}
\vspace{0.12in}
\centerline{\Large \hspace{0.15in}$\phi_2=0.2$\hspace*{1.5in}$\phi_2=0.5$}
\caption{ Construction of digitized two-dimensional 
realizations ($400\times400$ pixels) that putatively correspond to the target function
given by (\ref{E10})  for $\phi_2=0.2$ and $0.5$.\cite{Cu99,To02a} Here 
$a=32$ pixels and  $q=8\pi/a$. We now know that this function is not exactly realizable
because even though the construction technique matched (\ref{E10}) for almost
all $r$, it could not yield the necessary convex behavior in the vicinity of the origin.}
\label{Dinko}
\end{figure}

A generalization of the Debye random-medium function (\ref{debye})
that is nonmonotone and would
be characterized by short-range order is the following expression: \cite{Cu99}
\begin{equation}
f(r)= e^{-r/a}\;\frac{\sin(qr)}{qr},
\label{E10}
\end{equation}
where $q$ is an inverse length scale that controls oscillations 
in the term $\sin(qr)/(qr)$. 
The spectral function ${\tilde f}(k)$ of (\ref{E10})  in one, two, and three dimensions 
obeys the nonnegativity condition (\ref{necess}).  Interestingly, 
Torquato \cite{To02a} observed that although
(\ref{E10})  satisfies the upper bound of binary condition (\ref{chi-1}),
it does not necessarily satisfy the lower bound of (\ref{chi-1})
or, equivalently, the lower bound of (\ref{positive-2})
for all $\phi_1$, depending on the values of $a$ and $q$.
In other words, there are values of the infimum $f_{inf}$, which in this case
is a true global minimum, that violate the lower bound of (\ref{min-2}).
Let  $r_0$ be the radial distance at 
which $f(r)$ achieves its global minimum. The minima of $f(r)$ are solutions to the 
transcendental equation $q(a+r)\tan(qr)=q^2ar$. The extremum value
$qr_0$ can be shown to lie in the interval $[\pi, 3\pi/2)$ for arbitrary $a$ and $q$.
For example, for $aq=8\pi$, $r_0 \approx 5.671 a$ and
$f_{min}\approx -0.1818$ (see Fig. \ref{damped-sine}), 
For example, for $aq=8\pi$, $r_0 \approx 0.1772 a$ and
$f_{min}\approx -0.1818$ (see Fig. \ref{damped-sine}), 
and therefore,  according
to (\ref{intervals}), (\ref{E10}) is not realizable for the volume-fraction intervals
\begin{equation}
0 < \phi_i < 0.1538 \ldots \qquad  \mbox{and} \qquad 0.8461 \ldots <\phi_i <1.
\end{equation} 
Interestingly, two realizations of digitized two-dimensional two-phase media were
previously constructed \cite{Cu99,To02a} that putatively correspond to the scaled autocovariance function (\ref{E10}) 
for $\phi_2=0.2$ and $0.5$, respectively, and the aforementioned choice of $a$ and $q$
are shown in Figure~\ref{Dinko}. At $\phi_2=0.2$, the system resembles a {\it dilute particle 
suspension} with ``particle'' diameters of order $b$. At $\phi_2=0.5$, the resulting pattern  is {\it
labyrinthine} such that the characteristic sizes of the ``patches'' and ``walls'' are of  
order  $a$ and $2\pi/q$, respectively. For these sets of
parameters, all of the aforementioned necessary conditions
on the function are met, except for the triangular inequality.
Although (\ref{E10}) satisfies the negative slope condition 
(\ref{chi-slope}) at the origin, it only satisfies the convexity
condition (\ref{convex}) for $qa \le \sqrt{3}$, which we see
is violated in these instances, implying that the triangular inequality
must be violated. As it turned out, the construction
procedure matched the target function (\ref{E10}) for almost
all $r$, but it could not yield convex behavior in the vicinity of the origin.
Since the triangular inequality was not known at the time, it
was difficult to ascertain whether the slight discrepancy in
the curvature of the function at the origin was numerical imprecision.
We now know in retrospect that the construction technique was revealing
that a two-phase medium with a scaled autocovariance 
function (\ref{E10}) cannot be exactly realized,
which is a testament to the power of this method.

\section{Conclusions}

We have identified all of the known necessary conditions on the two-point correlation
function $S_2^{(i)}({\bf r})$ of statistically homogeneous two-phase media and have derived
a new but simple positivity condition that it must satisfy. Using these conditions,
we were able to ascertain the realizability of certain classes
of proposed correlation functions. In future work, it will be important
to identify other checkable necessary conditions. The stochastic optimization construction 
technique \cite{To02a} appears to be a very powerful numerical tool in guiding such
a search. Finally, we note that the analogous realizability problem for the
pair correlation function $g_2$ of point processes\cite{To02c,Cr03,Cos04,Ko05,Uc06} offers
many interesting challenges. It has recently been conjectured that
the known standard non-negativity conditions on $g_2$ are
sufficient to ensure the existence of point processes
at and above some sufficiently high space dimension\cite{To06a,To06b}.
Application of this conjecture implies the possibility
that the densest sphere packings in sufficiently high dimensions are
disordered rather than periodic, implying  the
existence of disordered classical ground states for some continuous
potentials. In future work, it would be interesting to investigate whether an analogous conjecture
applies to binary stochastic processes.

\noindent{\bf Acknowledgements}

It is a great pleasure and privilege for the author to contribute an article in 
a volume to honor the career of William B. Russel on the occasion of his 60th birthday. 
The author is grateful to John Quintanilla for valuable discussions and a critical
reading of this manuscript. This work was supported by the Office of Basic Energy Sciences, U. S.
Department of Energy, under Grant No. DE-FG02-04ER46108.


\end{document}